\documentclass[12pt]{article}
\usepackage{times}
\usepackage{epsf}
\usepackage{graphicx}

\def\beq{\begin{equation}}
\def\eeq{\end{equation}}
\def\bea{\begin{eqnarray}}
\def\eea{\end{eqnarray}}
\def\nnb{\nonumber}

\def\nnb{\nonumber}

\def\ba{\begin{array}}
\def\ea{\end{array}}
\def\bea{\begin{eqnarray}}
\def\eea{\end{eqnarray}}

\begin{document}
\title{PT/Non-PT Symmetric and Non-Hermitian P\"{o}schl-Teller-Like Solvable
Potentials via Nikiforov-Uvarov Method}
\author{\vspace{1cm}
         {\"Ozlem Ye\c{s}ilta\c{s}}
         \thanks{E-mail address: ozlem.yesiltas@yahoo.com.tr}
\\
{\small \sl  Turkish Atomic Energy Authority, Nuclear Fusion and Plasma Physics
Laboratory,}
\\{\small \sl Istanbul Road, 30 km Kazan 06983, Ankara, Turkey }}
\maketitle
\begin{abstract}
\noindent The solutions of trigonometric Scarf potential,
PT/non-PT-symmetric and non-Hermitian q-deformed hyperbolic Scarf
and Manning-Rosen potentials are obtained by solving the
Schr\"{o}dinger equation. The Nikiforov-Uvarov method is used to
obtain the real energy spectra and corresponding eigenfunctions.
\end{abstract}
~~~~~~~~~~{\small \sl PACS Ref: 03.65.Db, 03.65.Ge}


\newpage
\section{Introduction}
\indent The so called PT-symmetry of one dimensional quantum
mechanical potentials which  is the most recent symmetry concept is
defined as invariance under simultaneous space $P$ and time $T$
reflection appeared in quantum mechanics almost eight years ago [1].
Exact solution of the Schr\"{o}dinger equation for the potentials
which have complex spectrum are generally of interest. Potentials
admitting this symmetry are complex and non-hermitian, but an
interesting property of $PT$ symmetric quantum mechanics is that the
eigenvalue spectrum of these complex-valued Hamiltonians is real and
positive. It is also known that PT-symmetry does not necessarily
lead to completely real spectrum, and an extensive kind of
potentials with the real or complex form are being faced with in
various fields of physics. In particular, the spectrum of the
Hamiltonian is real if PT-symmetry is not spontaneously broken.
Recently, Mostafazadeh has generalized PT symmetry by
pseudo-Hermiticity  [2]. In fact, a Hamiltonian of this type is said
to be $\eta$- pseudo Hermitian if $H^{+}=\eta H \eta^{-1}$, where
$+$ denotes the operator of adjoint [3]. In [4] it was proposed a
new class of non-Hermitian Hamiltonians with real spectra which are
obtained using pseudo-symmetry. In the study of PT-invariant
potentials various techniques have been applied as variational
methods [5], numerical approaches [6], Fourier analysis [7],
semi-classical estimates [8], quantum field theory [9] and Lie group
theoretical approaches [10] and time dependent systems and
magnetohydrodynamics in plasma physics [11].
\\
\\
\noindent Recently, an alternative one called as Nikiforov-Uvarov
Method (NU-method) has been introduced for solving the
Schr\"{o}dinger equation. The well known potentials [12-16], Dirac
and Klein-Gordon equations for a Coulomb potential [17] by using the
NU-method are taken a part as applications of Schr\"{o}dinger
equation. Although NU method is a useful one that is successful to
solve Schr\"{o}dinger, Dirac, Klein -Gordon wave equations with
well-known central and non-central potentials, the method does'nt
work efficiently for all exactly solvable potential types [18]. The
origin of the problem is positive sign of derivative of $\tau$,
because the condition $\tau^{'}< 0$ helps to generate energy
eigenvalues and corresponding eigenfunctions. The NU method leads to
unacceptable energy values for a class of potentials such as
$PI(cosh(x)) $ that are studied by Levai and collaborators [10], due
to sign of $\tau^{'}> 0$  in the calculations. This difficulty is
improved in a recent work by an alternative method which is an
applicable scheme [18]. Therefore, the trigonometric Scarf,
q-deformed hyperbolic Scarf and Manning-Rosen potentials are in
solvable forms with the original NU approach. We write the
potentials in more general form with a $q$ deformation parameter
that may be used in describing the molecular interactions. The aim
of the present work is to obtain the energy eigenvalues and the
corresponding eigenfunctions of the P\"{o}schl-Teller-like
potentials as periodic Scarf which is in a trigonometric form,
q-deformed hyperbolic Scarf and Manning-Rosen potentials using the
NU-method within the framework of the PT-symmetric quantum
mechanics.
\\
\\
\noindent The organization of the paper is as follows. In Sec. II,
the Nikiforov-Uvarov method is briefly introduced. In Sec. III, IV
and V solutions of PT-/non-PT-symmetric and non-Hermitian forms of
the well-known  potentials are presented by using NU-method. The
results are discussed in Sec. VI.
\section{The Nikiforov-Uvarov Method}
\noindent The NU-method which has been developed by Nikiforov and Uvarov (NU-method)
is based on reducing the second order differential equations (ODEs) to a generalized
equation of hypergeometric type [15]. In this method, for a given $V(x)$, the
one-dimensional Schr\"{o}dinger equation is reduced to an equation which is
$\psi^{''}(s)+A(s)\psi^{'}(s)+B(s)\psi(s)=0$ type with an appropriate coordinate
transformation $x=x(s)$
\begin{eqnarray}
\psi^{''}(s)+\frac{\tilde{\tau}(s)}{\sigma(s)}\psi^{'}(s)+
\frac{\tilde{\sigma}(s)}{\sigma^{2}(s)}\psi(s)=0
\end{eqnarray}

\noindent where $A(s)=\frac{\tilde{\tau}(s)}{\sigma(s)}$ and
$B(s)=\frac{\tilde{\sigma}(s)}{\sigma^{2}(s)}$. In the $(1)$, $\sigma(s)$ and
$\tilde{\sigma} (s)$ are polynomials with at most second degree, and
$\tilde{\tau}(s)$ is a polynomial with at most first degree [15]. The wave function
is constructed as a multiple of two independent parts,
\begin{eqnarray}
\psi(s)=\phi(s) y(s),
\end{eqnarray}
\noindent and $(1)$ becomes [15]
\begin{eqnarray}
\sigma(s)y^{''}(s)+\tau(s)y^{'}(s)+\lambda y(s)=0,
\end{eqnarray}
\noindent where
\begin{eqnarray}
\sigma(s)=\pi(s)\frac{d}{ds}(ln \phi(s)),
\end{eqnarray}
\noindent and
\begin{eqnarray}
\tau(s)=\tilde{\tau}(s)+2\pi(s).
\end{eqnarray}

\noindent $\lambda$ is defined as
\begin{eqnarray}
\lambda_{n}+n\tau^{'}+\frac{[n(n-1)\sigma^{''}]}{2}=0, n=0,1,2,...
\end{eqnarray}

\noindent determine $\pi(s)$ and $\lambda$ by defining

\begin{eqnarray}
k=\lambda-\pi^{'}(s).
\end{eqnarray}

\noindent and $\pi(s)$ becomes

\begin{eqnarray}
\pi(s)=(\frac{\sigma^{'}-\tilde{\tau}}{2})\pm
\sqrt{(\frac{\sigma^{'}-\tilde{\tau}}{2})^{2}-\tilde{\sigma}+k\sigma}
\end{eqnarray}

\noindent The polynomial $\pi(s)$ with the parameter $s$ and prime
factors show the differentials at first degree. Since $\pi(s)$ has
to be a polynomial of degree at most one, in (8) the expression
under the square root must be the square of a polynomial of first
degree [15]. This is possible only if its discriminant is zero.
After defining $k$, one can obtain $\pi(s)$, $\tau(s)$,$\phi(s)$ and
$\lambda$. If we look at $(4)$ and the Rodrigues relation

\begin{eqnarray}
y_{n}(s)=\frac{C_{n}}{\rho(s)}\frac{d^{n}}{ds^{n}}[\sigma^{n}(s)\rho(s)],
\end{eqnarray}

\noindent where $C_{n}$ is normalization constant and the weight
function satisfy the relation as

\begin{eqnarray}
\frac{d}{ds}[\sigma(s)\rho(s)]=\tau(s)\rho(s).
\end{eqnarray}

\noindent where

\begin{eqnarray}
\frac{\phi^{'}(s)}{\phi(s)}=\frac{\pi(s)}{\sigma(s)}.
\end{eqnarray}

\section{The Trigonometric Scarf Potential}
\noindent The periodic Scarf potential which is in a trigonometric
form is given by [19]

\begin{eqnarray}
V(x)=-\frac{(\frac{1}{4}-s^{2})\pi^{2}}{2 m a^{2} sin^{2}(\frac{\pi x}{a})}
\end{eqnarray}

\noindent where, $a$ is the potential period. Let us write this
potential in a general form as

\begin{eqnarray}
V(x)=-\frac{A}{sin^{2}\alpha x}
\end{eqnarray}

\noindent In order to apply NU-method, one can write the Schr\"{o}dinger equation
with the generalized Scarf potential by using a new variable, $sin^{2}\alpha
x=1-s^{2}$

\begin{eqnarray}
\psi^{''}(s)-\frac{s}{1-s^{2}}\psi^{'}(s)+\frac{1}{(1-s^{2})^{2}}(-\varepsilon
s^{2}+\varepsilon+\beta)\psi(s)=0.
\end{eqnarray}

\noindent where $\frac{2mE}{\hbar^{2}\alpha^{2}}=\varepsilon$ and
$\frac{2mA}{\hbar^{2}\alpha^{2}}=\beta$. Substituting $\sigma(s)$,
$\tilde{\sigma}(s)$ and $\tilde{\tau}(s)$ in (8), one can obtain the
function of $\pi(s)$ as

\begin{eqnarray}
\pi(s)=-\frac{s}{2}\pm\sqrt{(\varepsilon-k+1/4)s^{2}-\varepsilon-\beta+k}
\end{eqnarray}

\noindent Due to NU-method, the expression in the square root must be the square of
a polynomial. Therefore, the new $\pi$ functions can be written for each $k$ as

\begin{eqnarray}
\pi(s)=-\frac{s}{2}\pm \left\{%
\begin{array}{ll}
    \sqrt{-\beta+\frac{1}{4}}, & \hbox{$k=\varepsilon+\frac{1}{4}$} \\
    s \sqrt{-\beta+\frac{1}{4}}, & \hbox{$k=\varepsilon+\beta$} \\
\end{array}%
\right.
\end{eqnarray}

\noindent After determining $k$ and $\pi$, we can write $\tau$ as,
\begin{eqnarray}
\tau(s)=-2s\left(1+\sqrt{-\beta+\frac{1}{4}}\right)
\end{eqnarray}

\noindent The correct value of $\pi$ is chosen such that the
function $\tau(s)$ given by (5) will have a negative derivative
[12-15]. So, one can obtain the energy eigenvalues as,

\begin{eqnarray}
E_{n}=\frac{\hbar^{2}\pi^{2}}{2ma^{2}}\left[(\frac{1}{2}+n)+\sqrt{\frac{1}{4}-\frac{2m
A}{\hbar^{2}\alpha^{2}}}\right]^{2}
\end{eqnarray}

\noindent These results of the bound state spectrum expression match
with the solutions given in [18]. Using (9-11), the wave function
can be written as,

\begin{eqnarray}
\psi(s)=(1-s^{2})^{\lambda/2}P^{\nu_{1},\nu_{2}}_{n}(1-s^{2}).
\end{eqnarray}

\noindent Here $P^{\nu_{1},\nu_{2}}_{n}(1-s^{2})$ stands for Jacobi polynomials and
$\nu_{1}=\nu_{2}=\frac{1}{2}+\lambda+n$,\,
$\lambda=\frac{1}{2}+\sqrt{-\beta+\frac{1}{4}}$.

\subsection{PT symmetric trigonometric Scarf potential}

In (13) we use $\alpha \rightarrow i \alpha$, then it turns into

\begin{eqnarray}
V(x)=\frac{A}{sinh^{2}\alpha x}
\end{eqnarray}

\noindent If we write this potential in the Schr\"{o}dinger \,\
equation and using a transformation as  $sinh^{2}\alpha x=s^{2}-1$,
the energy spectrum is obtained as

\begin{eqnarray}
E_{n}=-\frac{\hbar^{2}\alpha^{2}}{2m}\left(n+\frac{1}{2}-
\sqrt{\frac{2mA}{\hbar^{2}\alpha^{2}}+\frac{1}{4}}\right)^{2}.
\end{eqnarray}

\noindent In this case, in order to obtain the physical solutions,
there is a condition about the derivative of $\tau$ that is
explained in the last section as $\tau^{'} < 0$. \,\ Thus, the
condition $\sqrt{\frac{1}{4}+\frac{2mA}{\hbar^{2}\alpha^{2}}}< 1$ is
needed due to appropriate physical solutions.

\subsection{PT symmetric and q-deformed trigonometric Scarf potential}

\noindent If we use a mapping [21] as $x \rightarrow
x-\frac{1}{\alpha} ln\sqrt{q}$ in (20), it turns into a q-deformed
periodic Scarf potential as

\begin{eqnarray}
V_{q}(x)=\frac{A}{sinh_{q}^{2}\alpha x}
\end{eqnarray}

\noindent where $sinh_{q}x=\frac{1}{2}(e^{x}-qe^{-x})$,\,\  $q > 0$
and $q\rightarrow1, V_{q}(x)\rightarrow V(x)$ . We use
$sinh_{q}^{2}\alpha x=q(s^{2}-1)$ in the calculations and the energy
spectrum is

\begin{eqnarray}
E_{n}=-\frac{\hbar^{2}\alpha^{2}}{2m}\left(n+\frac{1}{2}-
\sqrt{\frac{2mA}{\hbar^{2}\alpha^{2}q}+\frac{1}{4}}\right)^{2}.
\end{eqnarray}

\noindent If the deformation parameter $q$ is taken as $q=1$, (23)
turns into the energy spectrum which is given in (21). The same
condition as $\sqrt{\frac{1}{4}+\frac{2mA}{\hbar^{2}\alpha^{2}}}< 1$
is valid in this potential calculations also.

\subsection{Non-PT symmetric, non-Hermitian and q-deformed trigonometric Scarf potential}

In this case, we write  $A\rightarrow A_{1}+ iA_{2}$ in (22), where
$A_{1},\,\ A_{2}$ and $\alpha$ are real, we choose $q\rightarrow
iq$, then

\begin{eqnarray}
E_{n}=-\frac{\hbar^{2}\alpha^{2}}{2m}\left(n+\frac{1}{2}-
\sqrt{-\frac{2m(iA_{1}-A_{2})}{\hbar^{2}\alpha^{2}q}+\frac{1}{4}}\right)^{2}.
\end{eqnarray}

\noindent As it is seen from (24), there is real energy spectrum in
case $A_{1}=0$.

\section{The q-Deformed Hyperbolic Scarf Potential}

\noindent The q-deformed hyperbolic Scarf potential is defined by
$(x>ln\sqrt{q})$ [20],

\begin{eqnarray}
V_{q}(x)=V_{0}+V_{1}coth^{2}_{q}\alpha x+V_{2} \frac{coth_{q}\alpha x}{sinh_{q}
\alpha x}
\end{eqnarray}

\noindent where $sinh_{q}x=\frac{1}{2}(e^{x}-qe^{-x})$ and
$cosh_{q}x=\frac{1}{2}(e^{x}+qe^{-x})$ and when $q\rightarrow 1$,
$V_{q}(x)\rightarrow V(x)$. The Schr\"{o}dinger equation with the
q-deformed Scarf potential by using a new variable $s=cosh_{q}\alpha
x$ is

\begin{eqnarray}
\psi_{q}^{''}(s)+\frac{s}{s^{2}-q}\psi_{q}^{'}(s)+
\frac{1}{(s^{2}-q)^{2}}((\varepsilon^{2}-\beta^{2})s^{2}-\gamma^{2}s-q
\varepsilon^{2})\psi_{q}(s)=0.
\end{eqnarray}

\noindent where $\varepsilon^{2}=\frac{2m (E-V_{0})}{\alpha^{2}\hbar^{2}}$,
$\beta^{2}=\frac{2m V_{1}}{\alpha^{2}\hbar^{2}q}$ and $\gamma^{2}=\frac{2m
V_{2}}{\alpha^{2}\hbar^{2} \sqrt{q}}$. The $\pi$ functions can be written for each
$k$ as,

\begin{eqnarray}
\pi_{q}(s)=\frac{s}{2}\pm \frac{1}{2} \left\{%
\begin{array}{ll}
    \zeta_{1}s+\zeta_{2},
     & \hbox{$k=-\frac{1}{8}+\varepsilon^{2}-\frac{\beta^{2}}{2}+\frac{\mu}{4}$} \\
    \zeta_{1}s-\zeta_{2} ,
    & \hbox{$k=-\frac{1}{8}+\varepsilon^{2}-\frac{\beta^{2}}{2}-\frac{\mu}{4}$} \\
\end{array}%
\right.
\end{eqnarray}

\noindent where,
$\zeta_{1}=\sqrt{\frac{1}{2}+2\beta^{2}+\frac{1}{8q}\mu}$,
$\zeta_{2}=\sqrt{\frac{1}{2}+2\beta^{2}-\frac{1}{8q}\mu}$,\,\,
$\mu=4q \sqrt{(4\beta^{2}+1)^{2}-\frac{16\gamma^{4}}{q}}$. With
appropriate choosing of $k$ and $\pi$, \,\  $\tau$ is written as

\begin{eqnarray}
\tau_{q}(s)=-(\zeta_{1}-2)s-\zeta_{2}
\end{eqnarray}

\noindent Thus, the energy eigenvalues are obtained as

\begin{eqnarray}
E_{n}=V_{1}+V_{0}- \frac{\alpha^{2}\hbar^{2}}{2m}\left[(n+\frac{1}{2})-
\frac{1}{2}\sqrt{\frac{1}{2}+\frac{4mV_{1}}{\alpha^{2}\hbar^{2}}+\frac{1}{2}\sqrt{(\frac{8m
V_{1}}{\alpha^{2}\hbar^{2}}+1)^{2}-\frac{64m^{2} V^{2}_{2}}{\alpha^{4}\hbar^{4}q
}}}\right]^{2}
\end{eqnarray}

\noindent which agree with the earlier results [20]. The wave
function can be obtained following the same way that is explained in
the section 3 as,

\begin{eqnarray}
\psi_{n}(s)=B_{n}(s^{2}-q)^{n-\frac{\nu_{1}}{4}-1}
e^{\nu_{2}tanh^{-1}\frac{s}{\sqrt{q}}}P^{\nu_{1},\nu_{2}}_{n}(s)
\end{eqnarray}

\noindent where $\nu_{1}=1-\sqrt{\frac{1}{2}+2\beta^{2}+\frac{\mu}{8q}}$ \,\,and\,\,
$\nu_{1}=\sqrt{\frac{1}{2}+2\beta^{2}-\frac{\mu}{8q}}$.

\subsection{PT symmetric and non-Hermitian q-deformed hyperbolic Scarf potential}
\noindent When $\alpha \Rightarrow i\alpha$ and
$V_{0},V_{1},V_{2},q$ are real, then the potential takes the form

\begin{eqnarray}
V(x)&=&V_{0}+V_{1}\frac{(1+q^{2})cos2\alpha x+4q-i(q^{2}-1)sin2\alpha
x}{(-1+q^{2})cos2\alpha x-4q-i(q^{2}-1)sin2\alpha x}+\nnb \\
&& \qquad \frac{2V_{2}}{\sqrt{q}}\frac{(1+q)cos\alpha x+i(-q+1)sin\alpha
x}{(-1+q^{2})cos2\alpha x-4q-i(q^{2}-1)sin2\alpha x}
\end{eqnarray}

\noindent where $i=\sqrt{-1}$. If we take $q=1$ in (31), it becomes

\begin{eqnarray}
V(x)=V_{0}+V_{1}\cos 2\alpha x+V_{2} \cos \alpha x
\end{eqnarray}

\noindent which is a type of Morse potential. The potential given in
(31) and such potentials are PT-symmetric and non-Hermitian but have
real spectra as

\begin{eqnarray}
E_{n}=V_{1}-V_{0}+ \frac{\alpha^{2}\hbar^{2}}{2m}\left[(n+\frac{1}{2})-
\frac{1}{2}\sqrt{\frac{1}{2}-\frac{4mV_{1}}{\alpha^{2}\hbar^{2}}+\frac{1}{2}\sqrt{(-\frac{8m
V_{1}}{\alpha^{2}\hbar^{2}}+1)^{2}-\frac{64m^{2} V^{2}_{2}}{\alpha^{4}\hbar^{4}q
}}}\right]^{2}
\end{eqnarray}

\subsection{Non-PT symmetric and non-Hermitian q-deformed hyperbolic Scarf potential}

\noindent In this case, if $V_{1}$ and  $V_{2}$ parameters are chosen as $V_{1}=
V_{1}+iV_{1}$, $V_{2}=V_{2}+ iV_{2}$ and $q\Rightarrow iq$ then the potential
becomes

\begin{eqnarray}
V_{q}(x)=V_{0}+(V_{1}+iV_{1})\frac{(qe^{-2\alpha x}-i)^{2}}{(qe^{-2\alpha
x}+i)^{2}}-\frac{(V_{2}+iV_{2})}{\sqrt{iq}}\frac{e^{-\alpha x}(1+iqe^{-2\alpha
x})}{(i+qe^{-2\alpha x})^{2}}
\end{eqnarray}

\noindent In this case, the energy spectrum is $(\hbar^{2}=2m=1)$

\begin{eqnarray}
E_{n}=V_{0}+i(V_{1}+iV_{1})+\alpha^{2}\left[(n+\frac{1}{2})-
\frac{1}{2}\sqrt{\frac{1}{2}+\frac{(2iV_{1}-V_{1})}{\alpha^{2}}+
\frac{1}{2}\sqrt{(\frac{\left(2iV_{1}-V_{1}\right)}{\alpha^{2}}+1)^{2}-
\frac{V^{2}_{2}}{q\alpha^{4}}}}\right]^{2}
\end{eqnarray}

\noindent As it can be seen from (34), in order to obtain real
energy spectrum, the parameters must be chosen as $Re(V_{1}=0)$ and
$Im(V_{2})=0$.

\section{The Manning-Rosen Potential}

\noindent The general form of the Manning-Rosen potential is given
by [20],

\begin{eqnarray}
V_{q}(x)=Acoth_{q}\alpha x+\frac{B}{sinh^{2}_{q}\alpha x}
\end{eqnarray}

\

\noindent where $q\rightarrow 1, V_{q}(x)\rightarrow V(x)$. The
potential is put into the Schr\"{o}dinger equation and the following
form is obtained with the new variable $s=e^{-2x}$ as

\begin{eqnarray}
\psi^{''}_{q}(s)+\frac{1-qs}{s(1-qs)}\psi^{'}_{q}(s)+
 \frac{1}{4s^{2}(1-qs)^{2}}(-\varepsilon (1-qs)^{2}-\beta (1-q^{2}s^{2})-\gamma
 s)\psi_{q}(s)=0.
\end{eqnarray}

\

\noindent where $\varepsilon=-\frac{2mE}{\hbar^{2}\alpha^{2}}$,
$\beta=\frac{2mA}{\hbar^{2}\alpha^{2}}$ and
$\gamma=\frac{8mB}{\hbar^{2}\alpha^{2}}$. Thus, one can easily get the energy
eigenvalues as,

\begin{eqnarray}
E_{n}=\frac{\hbar^{2}\alpha^{2}}{2m}\left[\frac{1}{4}\left(-(2n+1)+
\sqrt{1+\frac{\gamma}{q}}\right)^{2}+\frac{\beta^{2}}{4\left(-(2n+1)+
\sqrt{1+\frac{\gamma}{q}}\right)^{2}}\right]
\end{eqnarray}

\noindent which agree with the results [20]. The corresponding wave
function becomes,

\begin{eqnarray}
\psi_{n}(x)=e^{-2\sqrt{\varepsilon+\beta}x}(1-e^{-\nu
x})P^{(2\sqrt{\varepsilon+\beta},\,\,\nu-1)}_{n}(1-e^{-2x})
\end{eqnarray}

\

\noindent where $\nu=1-\sqrt{1+\frac{\gamma}{q}}$.

\subsection{PT symmetric and non-Hermitian Manning-Rosen potential}

In this case, if the parameter is chosen as $\alpha\Rightarrow i\alpha$, the
potential becomes

\begin{eqnarray}
V(x)=\frac{A((1-q^{2})cos2\alpha x+i(1+q^{2})sin2\alpha x)+4B}{(1+q^{2})cos2\alpha
x+i(1-q^{2})sin2\alpha x-2q}
\end{eqnarray}

\noindent In case of taking $q=1$, the potential becomes

\begin{eqnarray}
V(x)=\frac{i A \sin 2\alpha x+2B}{\cos 2\alpha x-1}
\end{eqnarray}

\noindent Hence the corresponding energy eigenvalues for the
potential given in (40) become

\begin{eqnarray}
E_{n}=-\frac{\hbar^{2}\alpha^{2}}{2m}\left[\frac{1}{4}\left(-(2n+1)+
\sqrt{1+\frac{\gamma}{q}}\right)^{2}+\frac{\beta^{2}}{4\left(-(2n+1)+
\sqrt{1+\frac{\gamma}{q}}\right)^{2}}\right]
\end{eqnarray}

\subsection{Non-PT symmetric and non-Hermitian Manning-Rosen potential}

\noindent In this case, if the parameters are chosen as $A=A_{1}+iA_{2}$, $B=B_{1}+i
B_{2}$ and $q\Rightarrow iq$, the potential turns into

\begin{eqnarray}
V_{q}(x)=i(A_{1}+iA_{2})\frac{1-q^{2}e^{-4\alpha x}}{(iqe^{-2\alpha x}-1)^{2}}+
 4(B_{1}+iB_{2})\frac{e^{-2\alpha x}}{(iqe^{-2\alpha x}-1)^{2}}
\end{eqnarray}

\noindent then the energy spectrum becomes ($2m=\hbar=1$) and
$\varepsilon=\frac{E}{4\alpha^{2}}$

\begin{eqnarray}
\varepsilon^{2}=-\frac{1}{16}\left(2n+1-i\sqrt{\frac{16b^{2}}{q}-1-8ia^{2}}\right)^{2}-
\frac{16ia^{2}}{\left(2n+1-i\sqrt{\frac{16b^{2}}{q}-1-8ia^{2}}\right)^{2}}
\end{eqnarray}

\noindent where $a^{2}=\frac{A_{1}+iA_{2}}{4\alpha^{2}}$ and
$b^{2}=\frac{B_{1}+iB_{2}}{4\alpha^{2}}$. As it is seen from (37),
it can be written $Re(a^{2})=0$,\,\, $Im(b^{2})=0$ and
$(\frac{16Re(b^{2})}{q}-1)<8\,\,Im(a^{2})$ due to obtaining real
energy spectrum.

\section{Conclusions}

\noindent  The PT-symmetric formulation have been extended to the
more general P\"{o}schl-Teller-like potentials as trigonometric
Scarf, q-deformed hyperbolic Scarf and Manning-Rosen potentials. The
Schr\"{o}dinger equation in one dimension have been solved for these
complex potentials by using Nikiforov-Uvarov method. It has been
shown that q-deformed Scarf and Manning-Rosen potentials have real
energy spectra without parameter restriction despite their
non-hermiticity. In addition, non-PT symmetric q-deformed Scarf and
Manning-Rosen potentials have real energy spectra in case there are
parameter restrictions. As an illustration, in Figure1-8, q-deformed
hyperbolic Scarf, PT/Non-PT symmetric q-deformed hyperbolic Scarf
potentials and Manning-Rosen, PT/Non-PT symmetric Manning-Rosen
potentials are plotted with different parameter values. As it is
seen from Figure2, there is a periodic behaviour of the PT-symmetric
q-deformed hyperbolic Scarf potential and there is a real energy
spectra due to unbroken PT symmetry. The Figure3 corresponds to
Non-PT symmetric potential and in case there are potential parameter
restrictions as $Re(V_{1})=0$ and $Im(V_{2})=0$, the energy spectra
is real. The real and imaginery parts of the PT symmetric
Manning-Rosen potentials are illustrated in Figure5-6, there is a
periodicity and unbroken PT symmetry as a result real energy
spectrum is obtained. In Figure7-8, Non-PT symmetric Manning-Rosen
potentials are illustrated and if $Re(a^{2})=0$,\,\, $Im(b^{2})=0$
and $(\frac{16Re(b^{2})}{q}-1)<8\,\,Im(a^{2})$, the real spectrum is
obtained. It is seen that, the potentials which are PT symmetric
shows a periodic behaviour and the energy spectrum is real without
parameter restrictions. Together with PT/Non-PT symmetric cases, it
has been pointed out that the applications of the miscellaneous
types of general P\"{o}schl-Teller-like potentials with real spectra
may be increased in different quantum systems.

\newpage
\renewcommand{\topfraction}{.99}
\renewcommand{\bottomfraction}{.99}
\renewcommand{\textfraction}{.01}
\renewcommand{\floatpagefraction}{.99}
\clearpage
\begin{figure}[htb]
\centering
\includegraphics{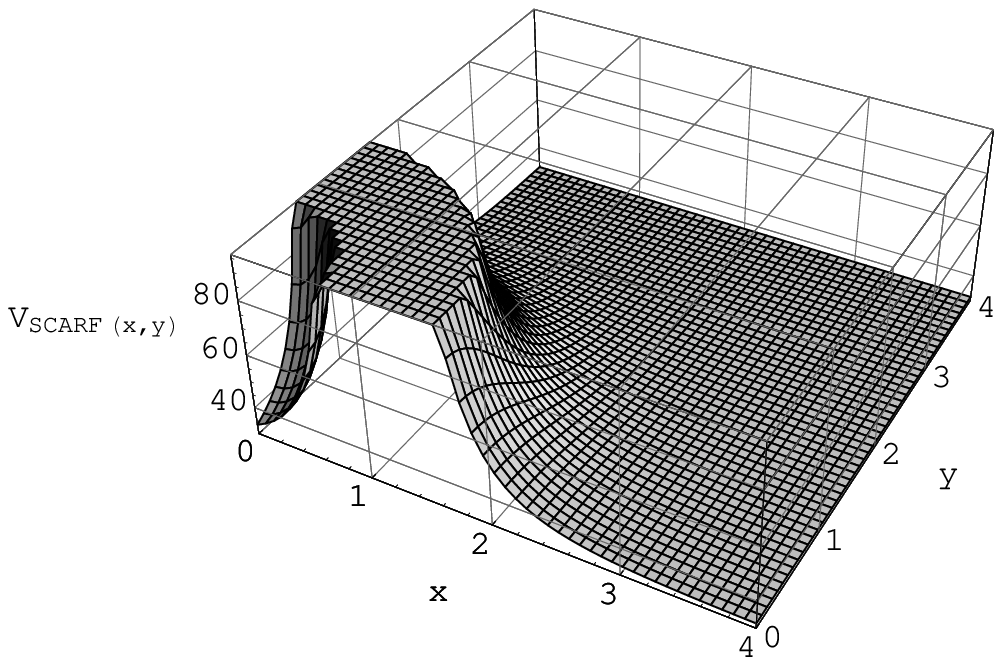}
\caption{The q-deformed Scarf potential;
$V_{0}=10,V_{1}=15,V_{2}=10,q=10,\alpha=1.$} \label{fig1}
\end{figure}

\begin{figure}[htb]
\centering
\includegraphics{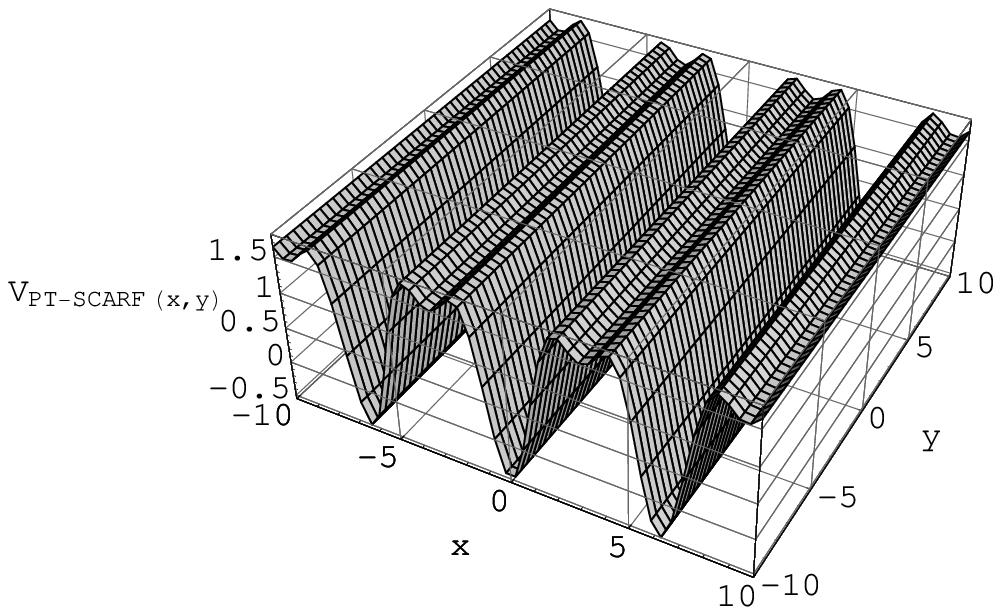}
\caption{PT-symmetric q-deformed Scarf
potential;$V_{0}=1,V_{1}=1,V_{2}=1,q=1,\alpha=1.$} \label{fig2}
\end{figure}

\begin{figure}[htb]
\centering
\includegraphics{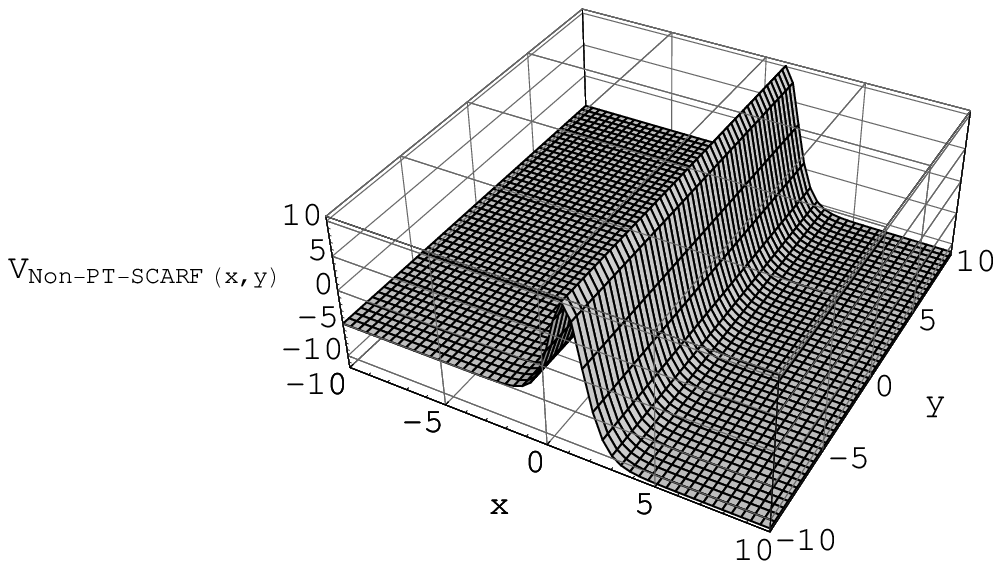}
\caption{Non-PT-symmetric q-deformed Scarf
potential;$V_{0}=10,V_{1}=15,V_{2}=10,q=10,\alpha=1.$} \label{fig3}
\end{figure}

\begin{figure}[htb]
\centering
\includegraphics{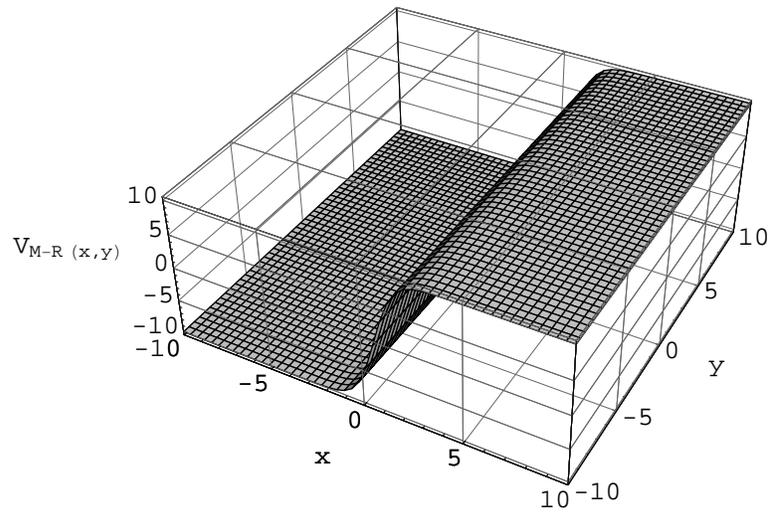}
\caption{The Manning-Rosen potential;$A=10,B=1,q=-4,\alpha=1.$}
\label{fig4}
\end{figure}

\begin{figure}[htb]
\centering
\includegraphics{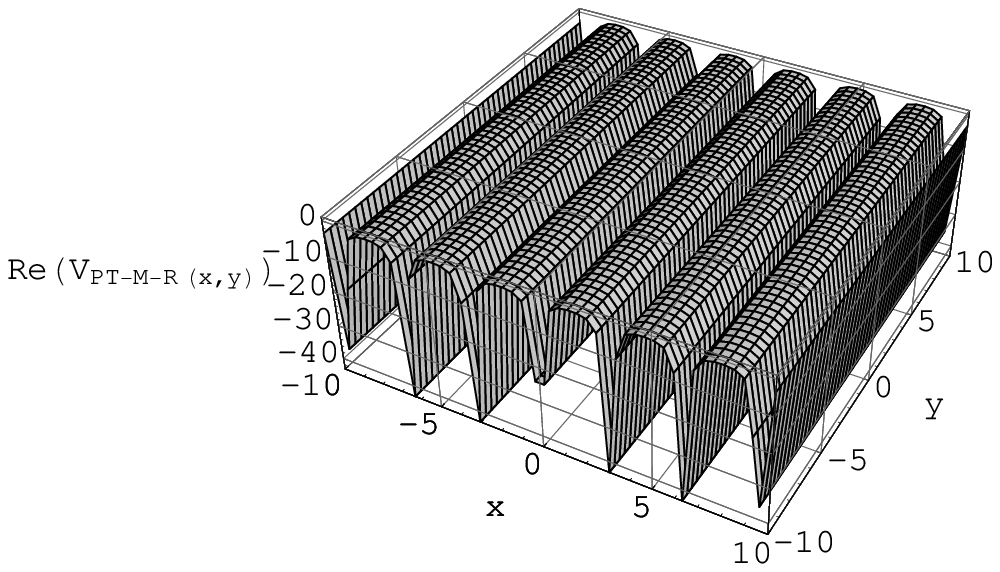}
\caption{The real part of the PT-symmetric Manning-Rosen
potential;$A=1,B=1,q=1,\alpha=1.$} \label{fig5}
\end{figure}

\begin{figure}[htb]
\centering
\includegraphics{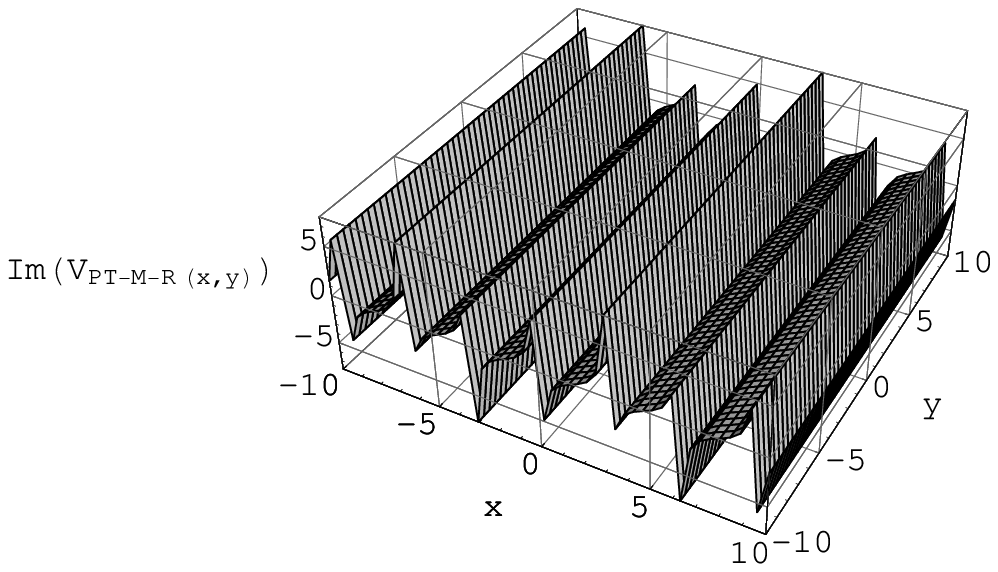}
\caption{The imaginer part of the PT-symmetric Manning-Rosen
potential;$A=1,B=1,q=1,\alpha=1.$} \label{fig6}
\end{figure}

\begin{figure}[htb]
\centering
\includegraphics{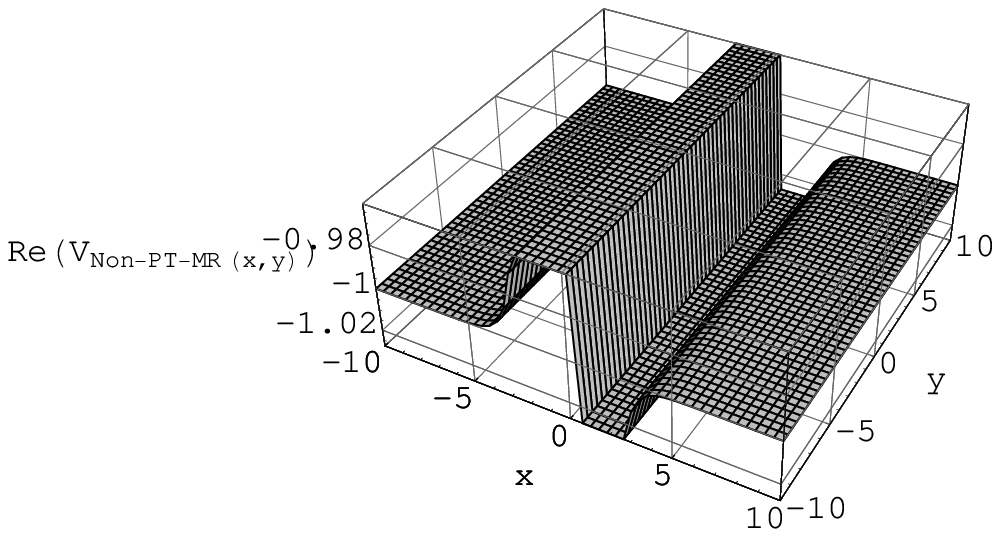}
\caption{The real part of the Non-PT-symmetric Manning-Rosen
potential;$A=1,B=1,q=1,\alpha=1.$} \label{fig7}
\end{figure}

\begin{figure}[htb]
\centering
\includegraphics{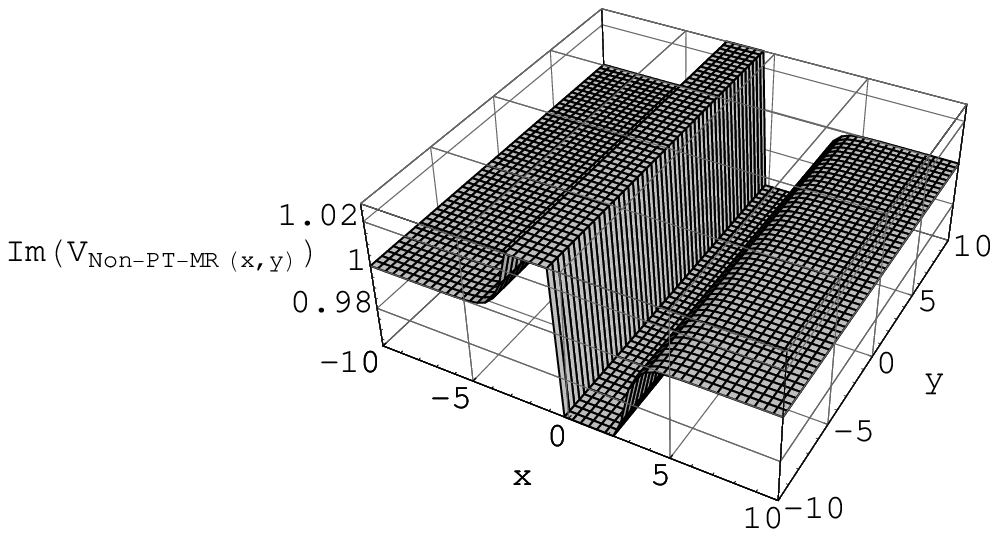}
\caption{The imaginer part of the Non-PT-symmetric Manning-Rosen
potential;$A=1,B=1,q=1,\alpha=1.$} \label{fig8}
\end{figure}

\end{document}